\documentclass[aps,prd,twocolumn,showpacs,superscriptaddress,groupedaddress,nofootinbib]{revtex4}  
\usepackage{graphicx}  
\usepackage{dcolumn}   
\usepackage{bm}        
\usepackage{amssymb}   
\usepackage{hyperref}
\usepackage{amsmath}
\begin{document}
\title{Prospects for indirect MeV Dark Matter detection with Gamma Rays\\ in light of Cosmic Microwave Background Constraints}

\author{Alma X. Gonz\'alez-Morales}
\email{alma.gonzalez@fisica.ugto.mx}
\affiliation{
Departamento de F\'isica, DCI, Campus Le\'on, Universidad de
Guanajuato, 37150, Le\'on, Guanajuato, M\'exico}
\affiliation{
Consejo Nacional de Ciencia y Tecnolog\'ia, Av. Insurgentes Sur 1582. Colonia Cr\'edito Constructor, Del. Benito Ju\'arez, C.P. 03940, M\'exico D.F. M\'exico}
\author{Stefano Profumo}
\email{profumo@ucsc.edu}
\affiliation{University of California, Santa Cruz, and Santa Cruz Institute  for Particle Physics, 1156 High St. Santa Cruz, CA 95060, United States of America
}
\author{Javier Reynoso-Cordova}
\email{reynosoj@fisica.ugto.mx}
\affiliation{
Departamento de F\'isica, DCI, Campus Le\'on, Universidad de
Guanajuato, 37150, Le\'on, Guanajuato, M\'exico}


\begin{abstract}
The self-annihilation of dark matter particles with mass in the MeV range can produce gamma rays via prompt or secondary radiation. The annihilation rate for such light dark matter particles is however tightly constrained by cosmic microwave background (CMB) data. Here we explore the possibility of discovering MeV dark matter annihilation with future MeV gamma-ray telescopes taking into account the latest and future CMB constraints. We study the optimal energy window as a function of the dominant annihilation final state. We consider both the (conservative) case of the dwarf spheroidal galaxy Draco and the (more optimistic) case of the Galactic center. We find that for certain channels, including those with one or two monochromatic photon(s) and one or two neutral pion(s), a detectable gamma-ray signal is possible for both targets under consideration, and compatible with CMB constraints. For other annihilation channels, however, including all leptonic annihilation channels and two charged pions, CMB data rule out any significant signal of dark matter annihilation at future MeV gamma-ray telescopes from dwarf galaxies, but possibly not for the Galactic center. 
\end{abstract}

\maketitle
%
%
\section{INTRODUCTION}\label{sec:introduction}

Dark matter is a key element in the current cosmological paradigm, the so-called $\Lambda$CDM concordance model of a cosmological constant plus cold dark matter. Such a picture is highly consistent with all latest cosmological observations, including those of the Cosmic Microwave background (CMB) and of galaxy distributions \cite{Ade:2015xua}. Little is known, however, about the dark matter fundamental properties as an elementary particle, and about whether or not the dark matter is coupled to the Standard Model other than via gravity. If dark matter particles annihilate (or decay -- we will not consider decay here, however) into Standard Model particles, it is possible to discover a non-gravitational signal through astrophysical observations. One way to search for this kind of signal is by looking for excess photon emission, typically in an energy range close to the dark matter particle mass, from dark-matter-rich targets such as the Galactic Center, local clusters of galaxies, Dwarf Spheroidal or local Milky-Way-like galaxies, where the expected signal-to-noise is often optimal  \cite{Profumo:2013yn}. 
An especially interesting case is when the dark matter annihilation events produce monochromatic photons in the final state, and the particle mass can be related to the energy of the expected photon lines (in the simplest case of a two-photon annihilation mode, the line approximately corresponds to the dark matter particle mass, for non-relativistic annihilating particles)  
\cite{Bergstrom:1997fj,Bringmann:2012ez,Goodman:2010qn,Rajaraman:2012db,Bergstrom:2012vd,Ibarra:2012dw,Yuksel:2007dr}.

The search for gamma rays as indirect probes of dark matter annihilation has been extensively pursued both theoretically and observationally; one of the most recent results is from the Fermi-LAT collaboration, and it covers the energy range $\sim$4.8 GeV up to $\sim$ 250 GeV \cite{Abdo:2010nc,Abdo:2010dk,Ackermann:2012qk}. The forthcoming GAMMA-400 space mission \cite{Galper:2014pua} is anticipated to launch at the beginning of 2020 and will search for gamma rays in the energy range from $\sim 100$ MeV up to $3$ TeV, thus overlapping Fermi's energy range. The energy range between $\sim 0.2$ MeV up to $\sim 100$ MeV is however still vastly untapped and largely unexplored \footnote{With the partial exception of the now defunct COMPTEL \cite{Lichti:1994bt} and EGRET \cite{Thompson:1993zz} telescopes, featuring however relatively low effective area and poor energy resolution.}, and potentially critical to search for dark matter with a mass in the few MeV to few hundred MeV range. Several proposed mission concepts have recently been discussed to deploy an MeV detector capable to eliminate this ``MeV gap'' \cite{Boddy:2015fsa}, including for example the e-ASTROGAM gamma-ray space mission \cite{Tatischeff:2016ykb} and many others such as GRIPS \cite{Greiner:2011ih}, PANGU \cite{Wu:2014tya}, ACT \cite{Boggs:2006mh}, and AdEPT \cite {Hunter:2013wla}.

The MeV range is a new exciting frontier for future indirect dark matter searches with gamma rays.   With this motivation in mind, in the present study we analyze  the projected capability of future MeV gamma-ray detectors in exploring MeV dark matter models, as a function of the dominant pair-annihilation finals states and compare with the tightest constraints on the allowed annihilation rate as a function of the particle mass stemming from  the induced distortions to the spectrum of the CMB \cite{Slatyer:2012yq,Slatyer:2015jla}.  We remain agnostic as to the specific UV realization of the particle models at hand, and assume that one of the kinematically allowed annihilation final states dominates. While previous works have presented detection limits for MeV dark matter \cite{Bartels:2017dpb,Boddy:2015efa}, as we illustrate in detail in Sec.\ref{sec:previous} our study extends and differs significantly in several aspects from previous analyses.

In this work we exclusively focus on s-wave annihilating dark matter. In the case of p-wave pair-annihilation the constraints from CMB are largely relaxed, as we discuss in section \ref{sec:thermalhistory}, but the prospects for gamma-ray detection are also not as promising as in the s-wave case. We compute the parameter space ranges on the ($m_\chi, \langle \sigma v \rangle_{\text{s-wave}}$) plane allowed by CMB for six different annihilation channels and we then proceed to compare those ranges with the values producing a $5 \sigma$ detection for some hypothetical detector specification, inspired by currently proposed experimental designs for future MeV detectors. 

This paper is organized as follows: In section \ref{sec:spectrum} we discuss the particle dark matter models and assumptions, and we present the photon spectrum for the different annihilation channels; In section \ref{sec:thermalhistory} we briefly discuss the thermal history of the Universe and how extra energy injection can alter the residual free-electron fraction after recombination, leading to distortions in the CMB Power Spectrum. In section \ref{sec:gammaraydetec} we discuss how we construct the hypothetical detector and  what energy range can enhance the detection for each channel, {\bf in section \ref{sec:previous} we present a brief discussion comparing previous works and our results}, and, finally, we conclude in section \ref{sec:discussion}.

\section{Gamma-rays from MeV Dark Matter Annihilation}
\label{sec:spectrum}
We consider dark matter masses in the range between the neutral pion mass ($\sim 135$ MeV) and 1 GeV. We remain agnostic about the underlying UV theory and about the spin of the dark matter particle; rather, we describe a given model realization by the triplet given by the dark matter particle mass $m_\chi$, the thermally-averaged zero-temperature pair annihilation cross section times relative velocity $\langle\sigma v\rangle$, and the dominant annihilation final state. For simplicity, we assume that, whenever kinematically open, the two-pion final state dominates over $n\pi$, $n>2$, and over final states involving heavier mesons, although this depends on the matching of the UV theory onto the light meson degrees of freedom \cite{Logan}. This is somewhat justified, however, because of phase-space suppression of the sub-dominant annihilation final states. 

With these assumptions, the two-body final states we consider in this work are:\\ 

\noindent ({\it i})  two photons, $\gamma \gamma$;\\
({\it ii}) photon and neutral pion, $\gamma \pi^0$, open for $\sqrt{s}>m_{\pi^0}$;\\
({\it iii} ) two neutral pions, $\pi^0 \pi^0$, for $\sqrt{s}>2m_{\pi^0}$; \\
({\it iv}) two charged pions, $\pi^+ \pi^-$, for $\sqrt{s} > 2m_{\pi^{\pm}}$;\\
({\it v} )  two charged leptons, $\bar{{l}}l$ (l=$e,\mu$) state, accessible for $\sqrt{s}>2m_{l}$.\\

Here $\sqrt{s}\simeq 2m_\chi$ is the  Mandelstam variable, $m_{\pi^0}$, $m_{e}$ and $m_{\mu}$ are the pion, electron, and  muon mass, respectively. We do not consider channels involving neutrinos since they do not affect the CMB nor do they produce (significant amounts of) photons. The $\gamma$-ray spectrum,$\frac{dN}{dE_\gamma}$, generated by the annihilation channels listed above are quite simple for the first three cases. For the $\gamma \gamma$ final-state the spectrum is a delta function centered at the dark matter particle mass,
\begin{equation}
\frac{dN}{dE_\gamma} = 2\delta({E_{\gamma}- m_\chi}). 
\label{eq:gamagama}
\end{equation}
The spectrum generated by the $\gamma \pi^0$ final state (see e.g. \cite{Bartels:2017dpb}) is a delta function for the prompt photon and a box-shaped spectrum for the subsequent decay of the $\pi^0$ into two photons, 
\begin{equation}
\frac{dN}{dE_\gamma} = \delta(E_\gamma-E_0) + \frac{2}{\Delta E}\theta(E_\gamma-E_-)\theta(E_+-E_\gamma), 
\label{eq:gamapion}
\end{equation}
where $E_0=m_\chi-\frac{m_{\pi^0}^2}{4m_\chi}$, $\Delta E=m_\chi-\frac{m_{\pi_0}^2}{4m_\chi}$, and $$E_\pm=\frac{m_\chi}{2}\left(\left(1+\frac{m_{\pi_0}^2}{4m_\chi}\right)\pm\left(1-\frac{m_{\pi_0}^2}{4m_\chi}\right)\right).$$

For the two neutral pions, $\pi^0 \pi^0$, we have a box-shaped spectrum
\begin{equation}
\frac{dN}{dE_{\gamma}}=\frac{4}{\Delta E}\theta(E_\gamma-E_-)\theta(E_+-E_\gamma),,
\label{eq:pionpion}
\end{equation}
where $\Delta E = E_+-E_-=\sqrt{\frac{s}{4} - m_{\pi^0}^2}$ and
$$
E_{\pm}=\frac{m_\chi}{2}\left(1\pm\sqrt{1-\frac{m_{\pi^0}^2}{m_\chi^2}}\right).
$$ For the  $\pi^+ \pi^-$ channel we used numerical results from Ref.~\cite{Logan}, the code provided computes the photons coming from a radiative process in the charged pion decay $\pi^{+(-)} \to \mu^{+(-)}\nu_\mu (\bar{\nu}_\mu) \to e^{+(-)} \nu_e (\bar{\nu}_e) \bar{\nu}_\mu (\nu_\mu)$ in a boosted frame, the energy of the charged pions comes from the reference frame of the dark matter annihilation. 

In the case of dark matter annihilating into leptons, the spectrum is  quite different since the photon final state comes from  radiative processes, and it is  approximately given by \citep{Bergstrom:2004cy,Bell:2008vx}
\begin{equation}
\frac{dN}{dy}\simeq\frac{\alpha}{\pi}\left(\frac{1-(1-y)^2}{y} \right)\left( \ln{\frac{s(1-y)}{m_{l}^2}-1}\right),
\label{eq:gamalep}
\end{equation}
where $y\equiv E_\gamma/m_\chi$ and the approximate leading-log formula applies for $m_\chi \gg m_\mu$. 

Notice that in this study we neglect secondary photon production \cite{Profumo:2010ya}. The most relevant process would be inverse Compton, but the typical energies for the up-scattered photons, even in the case of the most energetic and sufficiently dense photon background, typically starlight, for which $E_\gamma\sim1$ eV, would  be \cite{Profumo:2010ya,Colafrancesco:2006he,Colafrancesco:2005ji,Jeltema:2008hf} $$E^\prime_\gamma\sim\Gamma_e^2E_\gamma\sim\left(0.1\times \frac{m_\chi}{m_e}\right)^2E_\gamma\ll 1\ {\rm MeV},$$
where $\Gamma_e$ is the typical Lorentz factor of the $e^\pm$ produced in the dark matter annihilation event. Secondary photons thus largely fall outside the range of interest for future proposed MeV gamma-ray detectors.


\section{Thermal history and CMB constraints}\label{sec:thermalhistory}
The CMB is one of the most important observables on Cosmology. It has been measured with increasingly high precision, and the physics behind it is well understood. As a result, CMB data can be used to constrain dark matter models that inject electromagnetically interacting Standard Model particles, since those would alter the thermal history of the Universe. Specifically, dark matter self-annihilation injects energy in the intergalactic medium (IGM), with possible ionization and heating of the IGM gas, resulting in modifications to the  recombination process at redshifts $z\sim 1000$. Free electrons leftover after recombination interact with CMB photons and cause modifications to the CMB power spectrum, which, in turn can be constrained with current CMB data. 

The energy per unit volume per unit time successfully injected in the IGM by dark matter particle pair-annihilation is usually cast as ~\cite{Chen:2003gz}:
\begin{equation}
\frac{dE}{dtdV}=\rho_c^2 c^2 \Omega_\chi^2(1+z)^6 P_{\text{ann}}(z),
\label{eq:Pann}
\end{equation}
where $\rho_c$ is the critical density of the Universe, $\Omega_\chi$ is the dark matter density and the annihilation parameter, 
\begin{equation}
P_{\text{ann}} \equiv f(z)\frac{\langle \sigma v \rangle}{m_\chi}
\end{equation}
is given in terms of the dark matter particle mass, $m_\chi$, the thermally averaged cross section $\langle \sigma v \rangle$, and a redshift-dependent efficiency function $f(z)$. Equation \eqref{eq:Pann} is coupled to the evolution of the free-electron fraction and medium temperature, so one has to solve both and include the result in the CMB fluctuations analysis. The standard methodology of solving these equations in presence of dark matter annihilations is extensively described in Ref~\cite{Chen:2003gz}, and is implemented in Boltzmann codes as CLASS \cite{Lesgourgues:2011re}.

In this work, we use the current constrains to the s-wave dark matter annihilation, given by the latest Planck constraints \cite{Ade:2015xua},
\begin{equation}
P_{\text{ann}}<4.1\times 10^{-28} \text{cm}^3 \text{s}^{-1} \text{GeV}^{-1},
\label{eq:planck_constraint}
\end{equation}
which, by means of equation \eqref{eq:Pann}, translates into an excluded region in the mass \emph{vs} cross-section plane. However, the annihilation probability $P_{\text{ann}}$ is in principle  a redshift-dependent quantity through the efficiency function. Fortunately, it has been demonstrated \cite{Slatyer:2015jla} that one can use an effective redshift-independent efficiency function $f_\text{eff}$; the authors of Ref.~\cite{Slatyer:2015jla} have proved that by making this change, the CMB power spectrum is altered in the same way as if one were including a redshift-dependent efficiency function. Following  reference \cite{Slatyer:2015jla}, the $f_{\text{eff}}$ is computed as:
\begin{equation}
f_{\text{eff}}=\frac{1}{2m_\chi}\int_{0}^{m_\chi} E dE \left( f_{\text{eff}}^{\gamma}(E)\frac{dN}{dE_{\gamma}} + 2f_{\text{eff}}^{e^+}(E)\frac{dN}{dE_{e^+}} \right), 
\label{eq:feff}
\end{equation}
where the functions $f_{\text{eff}}^{(\gamma)(e^+)}$ are  provided in Ref.~\cite{Slatyer:2015jla}. 

\begin{figure}[t!]
\centering
\includegraphics[width=1\linewidth]{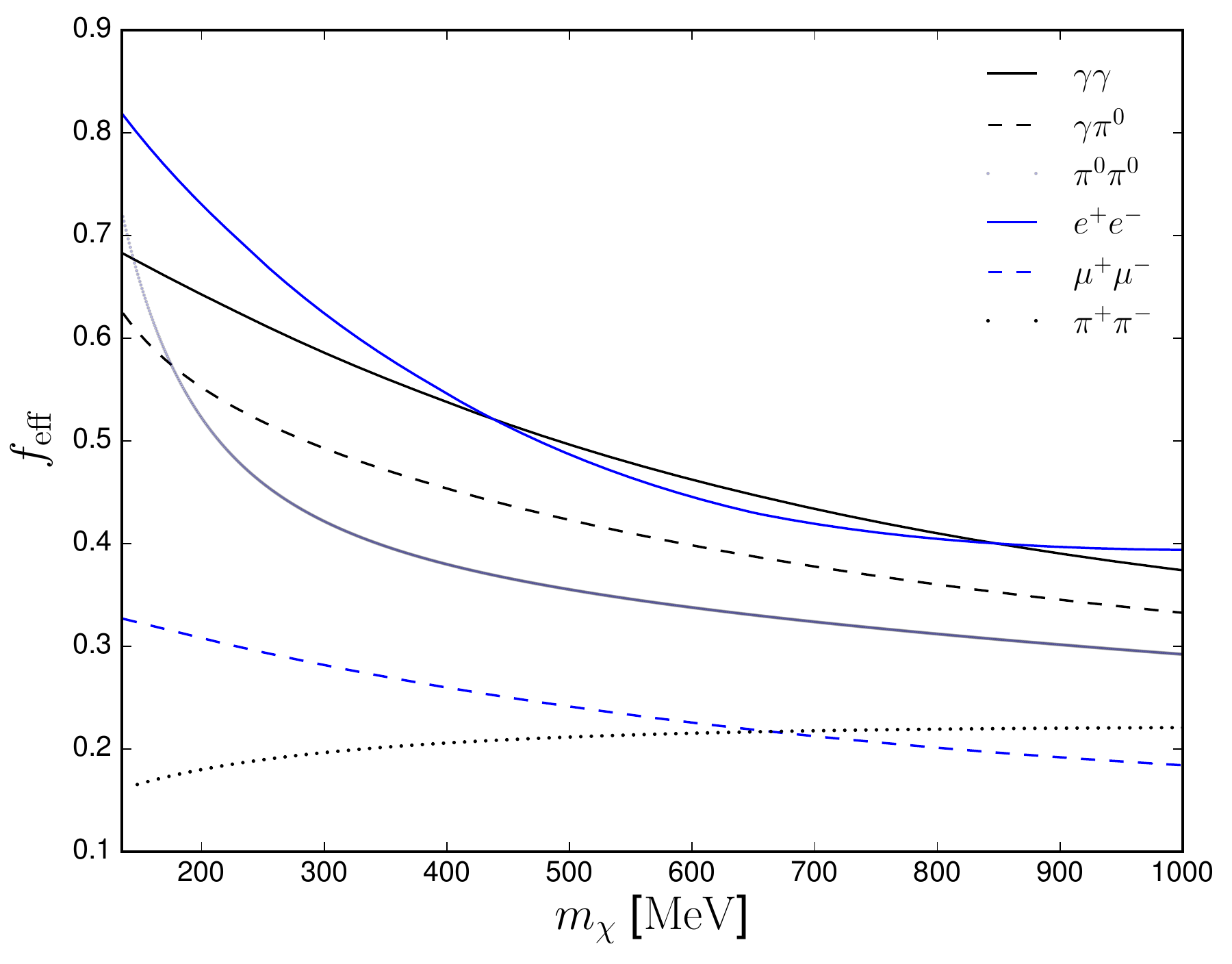}
\caption{Energy-injecting efficiency functions $f_{\rm eff}$ for the six different dark matter annihilation channels considered in this work.}
\label{Fig:feff}
\end{figure}

One therefore exclusively needs to know the injected photon and electron-positron spectrum for a given annihilation final state to compute the effective efficiency function and apply the Planck constraints, Eq.~\eqref{eq:planck_constraint}, for each annihilation channel. For the cases of dark matter annihilating into $\gamma \gamma$, $\gamma \pi^0$, $\pi^0 \pi^0$ and $e^+ e^-$ this is straightforward using the spectra presented in Section \ref{sec:spectrum}, Eq.~\eqref{eq:gamagama}, \eqref{eq:gamapion} and \eqref{eq:pionpion}. For the  $e^+ e^-$ case we need to add a delta-like function centered at the dark matter particle mass in addition to the final state radiation spectrum, while for the muon pair case, in addition to the photon spectrum from final state radiation off the muons, Eq.~\eqref{eq:gamalep}, we use the secondary electron-positron spectrum fit given in Ref.~\cite{Cirelli:2010xx}, valid since we are in the range $m_\chi > m_\mu$. Finally, for the case of charged pions, the electron-positron spectrum was computed following the results of Ref.~ \cite{chargedpionsspectrumMilford,Medhi2016,Culverhouse:2006kb}. The effective functions $f_{\rm eff}$ for all these channels as a function of the dark matter particle mass are presented in Fig.~\ref{Fig:feff} \footnote{A python code to compute the efficiency functions is available at https://github.com/JavierReynoso/feff.git.}.

With all these ingredients in hand, one can set constrains on the parameter space ($\langle \sigma v \rangle$, $m_\chi$) through:
\begin{equation}
\langle \sigma v \rangle < \frac{m_\chi}{f_{\text{eff}}}P_{\text{ann}}.
\label{eq:cross_section_planck}
\end{equation}
Such constraints correspond to the  solid lines in Fig. \ref{Fig:cases1} and in the following figures, which will be discussed in detail below.

Thus far we have only  discussed  CMB constraints  for an s-wave annihilating dark matter cross-section, but  if we allow the thermally averaged cross section to be velocity dependent, $\langle \sigma v \rangle \propto v^2$,  CMB constraints relax very significantly.
Specifically, the injected energy due to p-wave annihilating dark matter is
\begin{equation}
\frac{dE}{dVdt}= c^2 \Omega_\chi \rho_c^2(1+z)^6 f(z)\frac{\langle \sigma v \rangle_p}{m_\chi},
\label{eq:dEpwave}
\end{equation}
where 
\begin{equation}
\langle \sigma v \rangle_p = ( \sigma v )_\text{ref} \frac{\langle \upsilon \rangle^2}{\langle \upsilon \rangle_\text{ref}^2}=( \sigma v )_\text{ref}\frac{(1+z)^2}{(1+z_\text{ref})^2}, 
\label{eq:sigmav_pwave}
\end{equation}
note that  $\langle \sigma v \rangle_p \propto T_\chi$ \cite{Profumo:2013yn}.
Equation \eqref{eq:dEpwave} results in a suppression on the energy injection and thus will not alter the thermal history until low redshift. At the redshifts where dark matter contributes one must also consider the clumping effect due to the formation of dark matter halos~\cite{Liu:2016cnk}. In addition, to compute the $z_\text{ref}$ one must know the temperature of kinetic decoupling ($T_{\text{kd}}$), which is model dependent (see e.g. Ref.~\cite{Profumo:2006bv,Cornell:2013rza,Cornell:2012tb}). Given that constraints from CMB for p-wave annihilation are both weak and model-dependent, and that, moreover, the corresponding detectability of a gamma-ray signal is highly dependent on the velocity distribution in the target dark matter halo, in this work we exclusively focus on s-wave annihilators. Limits on p-wave annihilating dark matter from CMB for larger dark matter masses in standard WIMP scenarios have been presented in  \cite{Liu:2016cnk,Diamanti:2013bia,Choquette:2016xsw}.


\begin{figure}[ht!]
\centering
\includegraphics[width=1\linewidth]{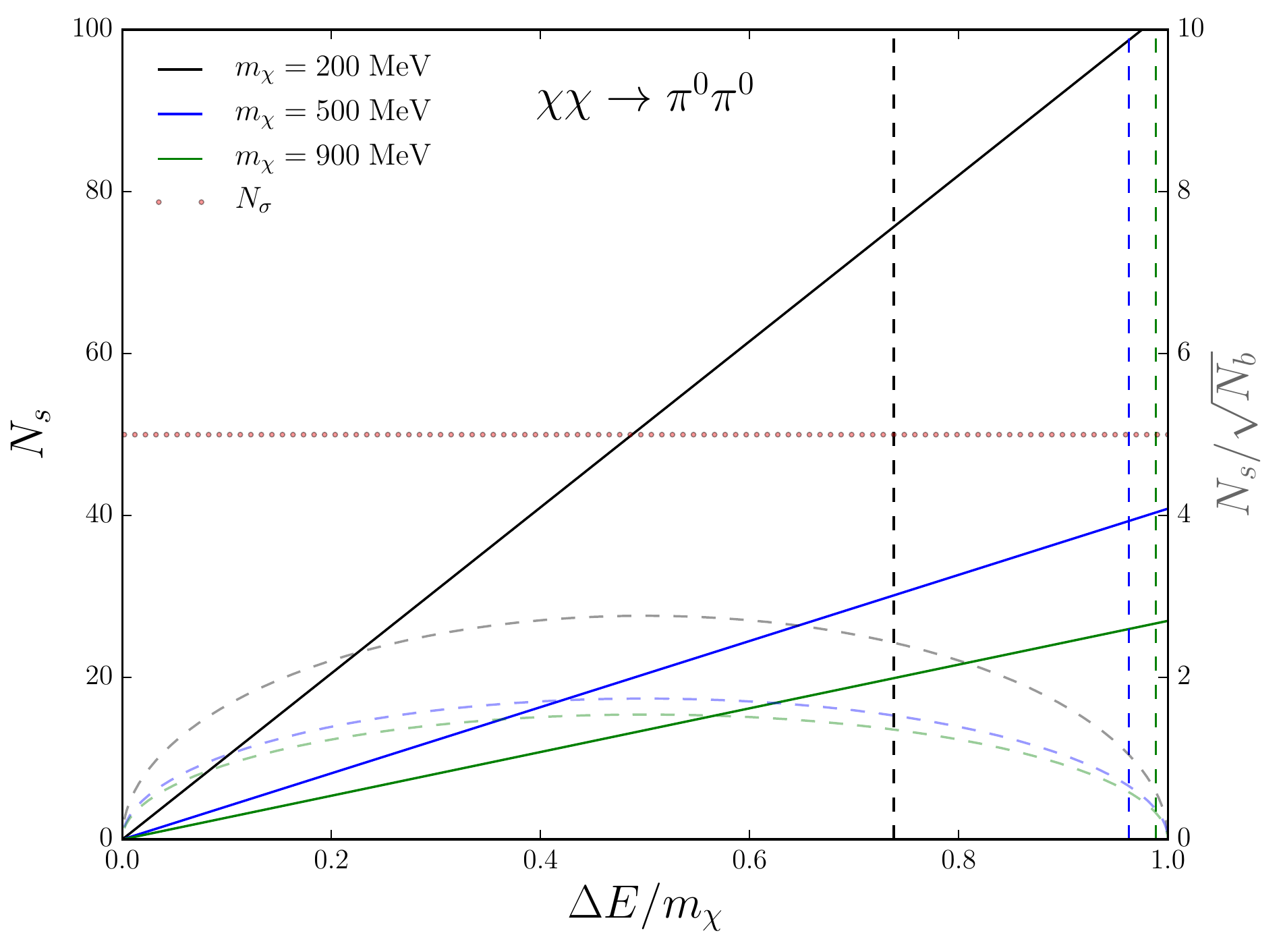}
\caption{Results of the analysis on the integration range in the photon spectrum and diffuse-background flux. We present the number of event photons (left axis, solid lines) in the energy range $\Delta E/m_\chi$. On the right axis (dashed lines) we show the corresponding signal-to-noise ratio $N_s / \sqrt{N_b}$ ($\# \sigma$) vs $\Delta E/m_\chi$. The vertical dashed lines represent the maximum energy-range possible for a certain mass in the case of neutral pions.} 
\label{Fig:NeutralPions}
\end{figure}

\begin{figure*}[!ht]
\includegraphics[width=\textwidth]{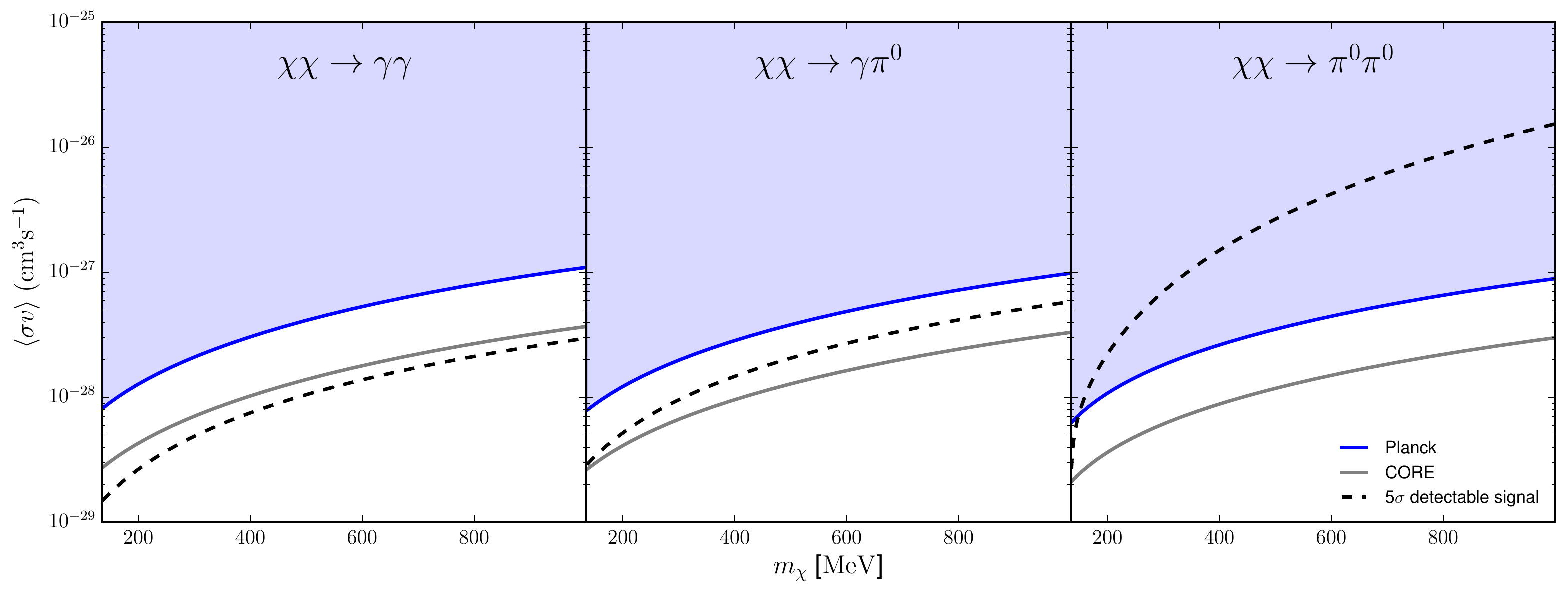}
\caption{Comparison on the values of $\langle \sigma v \rangle$ needed for a $5\sigma$ detection in the hypothetical gamma-ray detector described in the text from dark matter annihilation in the dSph Draco and the current constraints from Planck. The dashed lines represent the $\langle \sigma v \rangle$ needed for a $5\sigma$ detection while the solid lines represent the Planck constraints. The gray line is the projection constraint from COrE+, TEP. The yellow-colored region is ruled out by the Planck constraints.} 
\label{Fig:cases1}
\end{figure*}

\section{Gamma-ray Detection}
\label{sec:gammaraydetec}
What is the optimal energy window to search for gamma rays from MeV-scale dark matter particles? The question involves at the same time selecting energy windows and targets with a large enough signal to collect a significant number of signal photon events, and on optimizing the signal-to-noise ratio. 
The photon flux from dark matter pair annihilation is given by:
\begin{equation}
\phi =  J(\Delta\Omega)\cdot \frac{1}{4\pi}\frac{\langle \sigma v\rangle}{2m_\chi^2}\int dE \frac{dN}{dE_\gamma},
\end{equation}
where $J$ is the astrophysical ``J-factor'', the line of sight integral of the dark matter density squared integrated over an angular window subtending a solid angle $\Delta\Omega$.  In this work we focus on the dwarf spheroidal galaxy Draco, with a a J-factor of  $\text{log}_{10}(J/\text{GeV}^2\text{cm}^{-5})=19.05^{+0.22}_{-0.21}$ \cite{Boddy:2016fds}, and on the Galactic center, in which case we will consider a broad range of possible values for  J-factor, $\text{log}_{10}(J/\text{GeV}^2\text{cm}^{-5}) \approx {21}--{23}$.  The actual values were computed for the compilations of profiles made in \cite{Gammaldi:2016uhg}. Such compilation includes profiles found in state-of-the art $N$-body + hydrodynamical simulations of Milky Way like galaxies, namely the MOLL \cite{Mollitor:2014ara}, EAGLE \cite{Calore:2015oya}, ERIS \cite{Guedes:2011ux}, GARROTXA (GARR)\cite{Roca-Fabrega:2015gma}, and it also includes a DM-only profile (EVANS )\cite{PhysRevD.69.123501}; all these profiles satisfy the constrain of the DM abundance at the Solar System. This should cover the large uncertainty associated to the inner density profile of the Milky Way. As for the solid angle, in the case of Draco we take $\Delta\Omega=1.6\times 10^{-3}$ sr, corresponding to the angular area subtended in the sky, while for the Galactic center we use the solid angle corresponding to an aperture of $1^\circ$ ($\Delta\Omega=9.5\times 10^{-4}$ sr). This corresponds to the angular resolution planned for e-ASTROGAM over most of the energy range of interest here \cite{Tatischeff:2016ykb}; setting the angular region of interest to the angular resolution is intended to maximize the signal-to-noise for dark matter annihilation in the Galactic center.

The number of photons coming from a given target with a given J-factor is is given by:
\begin{equation}
N_s = A_{\text{eff}}\cdot T_{\text{obs}}\cdot  \phi,
\end{equation}
where  $A_{\text{eff}}$ is the effective area of the detector, $T_{\text{obs}}$ is the time of observation. The total number of collected signal photons must be large enough so that the corresponding signal-to-noise ratio yields a statistically significant detection. We here assume that a number of signal photon $N_s \sim N_{\sigma}\sqrt{N_b}$, where $N_b$ is the number of background photons that corresponds to a detection of statistical significance $N_{\sigma}$. 
As mentioned in the introduction, our main goal is to explore the plausibility of a dark matter signal using future MeV gamma-ray telescopes. To perform this analysis we considered a hypothetical detector with specifications similar to the proposed ASTROGAM detector \cite{Tatischeff:2016ykb}; specifically, we assume an effective area of $A_{\text{eff}}=500\ \text{cm}^2$ and an observation time $T_{\text{obs}}=1 \hspace{1mm}\text{year}$. Using these numbers and requiring a 5$\sigma$ detection, $N_{\sigma}=5$, we can derive an expression for $\langle \sigma v \rangle$ in terms of the mass that would guarantee this detection.
\begin{equation}
\langle \sigma v\rangle > 10\sqrt{N_b} \frac{1}{\int_{E_{\text{min}}}^{E_{\text{max}}}dE \frac{dN}{dE_\gamma}}\frac{4\pi}{A_{\text{eff}}T_{\text{obs}}J}m_\chi^2.
\label{eq:cross_section_det}
\end{equation}
\noindent To fully compute the values for $\langle \sigma v \rangle$ that can satisfy this, we must know the number of background photons $N_b$ and the integrated gamma-ray spectrum coming from dark matter annihilations. On one hand, we have that the number of background photons $N_b$ is proportional to the integrated background diffuse gamma-ray spectrum, which we assume, following Ref.~\cite{Boddy:2016fds}, to be given by:
\begin{equation}
\frac{d\phi}{d\Omega dE} = (2.74)\times10^{-3}\left(\frac{\text{MeV}}{E}\right)^{-2.0} \text{cm}^{-2}\text{s}^{-1}\text{sr}^{-1}\text{MeV}^{-1},
\end{equation}
as obtained from a fit to data from COMPTEL \cite{comptel} and EGRET \cite{Strong:2004ry}. For the case of the Galactic center, we utilize the so-called region A of the analysis in Ref.~\cite{Strong:2004ry}, giving a level of diffuse emission with a similar spectrum but a factor about 4 times larger. 
Now, the challenge is to find the optimal integration range for the gamma-ray signal spectrum and background: picking an arbitrarily large integration range may be best for some cases but decreases the detection line in others.  We thus proceeded to optimize the search strategy by picking the best integration energy range that gives a maximum $N_s/\sqrt{N_b}$ for each channel, assuming a lower limit of $m_\chi - \Delta E$ and an upper limit of $m_\chi$ and analyzing the results as a function of $\Delta E$ to select the $\Delta E$ that maximizes the signal-to-noise ratio,
\begin{equation}
\frac{N_s}{\sqrt{N_b}} \propto f(\langle \sigma v \rangle, m_\chi)
\end{equation}
The goal of this analysis is, given a certain $m_\chi$ and $\langle \sigma v \rangle$, to calculate which $\Delta E$ maximizes $N_s/\sqrt{N_b}$ while still giving enough photon events $N_s$ (we aimed for a minimum of $N_s\sim 20$).  To do so, we explored the signal to noise as a function of $\Delta E$ for a few different representative sample mass cases. 
While for the first three final states listed in Sec. \ref{sec:spectrum} analytical integration is straightforward,  for the lepton and charged pion cases we resorted to numerical integration. 
We found that for the cases of dark matter annihilating into $\gamma \gamma$ and $\gamma \pi^0$ the optimal range corresponds to the {\em smallest possible energy window}, which we take to be as low as the energy resolution of the detector, $\Delta E/E \sim 1\%$, again having in mind ASTROGAM \cite{Tatischeff:2016ykb} which is designed to achieve this energy resolution. For the leptons and charged pion cases, we found  that a $5 \sigma$ detection is not possible since the number of photons in this energy range of ($m_{\pi^0} < E < 1$ GeV) are not enough to even have the required event photons with maximal cross sections corresponding to the CMB limits. Nevertheless we picked a $\Delta E/E \sim 0.9$, which is the value that maximizes the signal to noise.  

\begin{figure*}
\centering
\includegraphics[width=\textwidth]{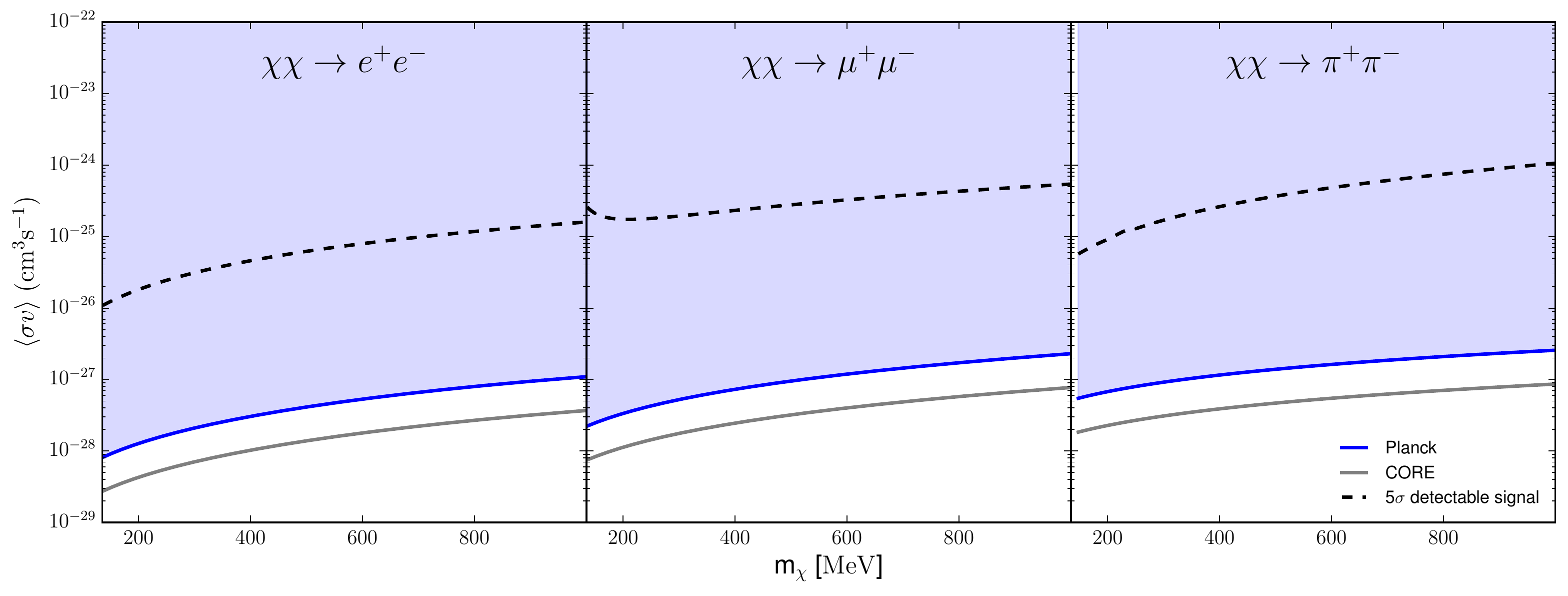}
\caption{As in Fig.~\ref{Fig:cases1}, but for the $e^+e^-$, $\mu+\mu^-$ and $\pi^+\pi^-$ final states.}
\label{Fig:cases2}
\end{figure*}

The case of dark matter annihilating into neutral pions is the most interesting one. Given the specific shape of the gamma-ray spectrum, there is a maximum possible integration range, that means our $\Delta E$ cannot be larger that ($\sqrt{s/4 - m_{\pi^0}^2}$), to the expense of only integrating additional background; this led us to consider a mass-dependent $\Delta E$ for each value of the dark matter mass that lies across the allowed integration range.  The result of this analysis for the annihilation into neutral pions is presented in Fig. \ref{Fig:NeutralPions}, where we present the number of event photons (left axis) in the energy range $\Delta E/m_\chi$. The right axis, corresponding to the dashed lines, shows the corresponding signal to noise ratio $N_s / \sqrt{N_b}$ ($N_{\sigma}$), a proxy of the statistical significance. The vertical dashed lines represent the maximum energy-range possible for a certain mass in the case of neutral pions. We only show three specific masses, the low and high limits, and one in between. As we mentioned before, we chose the maximum possible integration range for the $\pi^0 \pi^0$ case, in Fig. \ref{Fig:NeutralPions} the maximum clearly occurs at $\Delta E /E \sim 0.5$, but for masses near the pion mass, the vertical line, which is the maximum possible integration range, would be positioned before the 0.5 value which lead us to think that a fixed value for $\Delta E/E$ would give an over estimation in some cases. Therefore, and for simplicity, and even though this $\Delta E$ does not always maximizes the detection,  we will use the maximum possible $\Delta E$ for each mass given by ($\sqrt{s/4 - m_{\pi^0}^2}$).

Having performed the optimization analysis described above, we proceeded to compare the values of $\langle \sigma v \rangle$ we need for a $5\sigma$ detection with the current s-wave Planck constraints for the different final state channels. In Fig. \ref{Fig:cases1} we present the case of dark matter annihilating into $\gamma \gamma$, $\gamma \pi^0$ and $\pi^0 \pi^0$, for the Draco dSph. For all three cases, we find that there is a mass range allowed by Planck constraints where a signal can be detected, although for the case of neutral pions that is limited to masses very close to the pion threshold.
In Fig. \ref{Fig:cases2} we present the cases of dark matter annihilating into $e^+ e^-$, $\mu^+ \mu^-$ and $\pi^+ \pi^-$. The figure illustrates how for charged particles no MeV gamma-ray signal is possible from the dSph Draco due to Planck constraints. In addition, future CMB limits are shown, gray lines, indicating the projected constraints from COrE+ (TEP specification), at the level of $P_{\rm{ann}}< 1.38 \times 10^{-28} \rm{cm}^{3} \rm{s}^{-1} \rm{GeV}^{-1}$ \cite{DiValentino:2016foa}.

\begin{figure*}
\centering
\includegraphics[width=\textwidth]{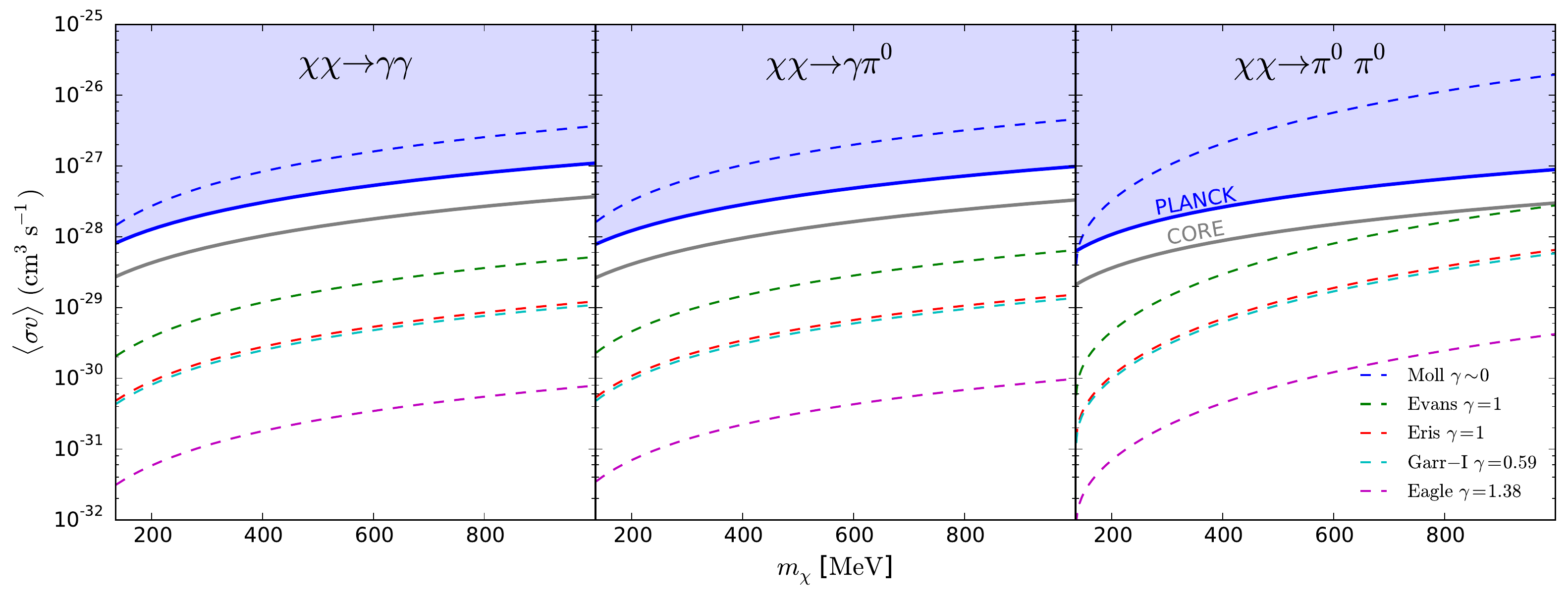}
\caption{ Comparison on the values of $\langle \sigma v \rangle$ needed for a $5\sigma$ detection in the hypothetical gamma-ray detector described in the text, from dark matter annihilation in Galactic Center, for the $\gamma \gamma$, $\gamma \pi^0$ and $\pi^0 \pi^0$ final states. The dashed lines correspond to the different density profiles used to compute the "J-factor", the label indicates the name of the simulation and the inner slope of the density profile (see the text for more details). The blue region is ruled out by the PLANCK constraints (blue line), while the gray line is the projection constraint from COrE+, TEP.}
\label{Fig:cases3}
\end{figure*}

\begin{figure*}
\centering
\includegraphics[width=\textwidth]{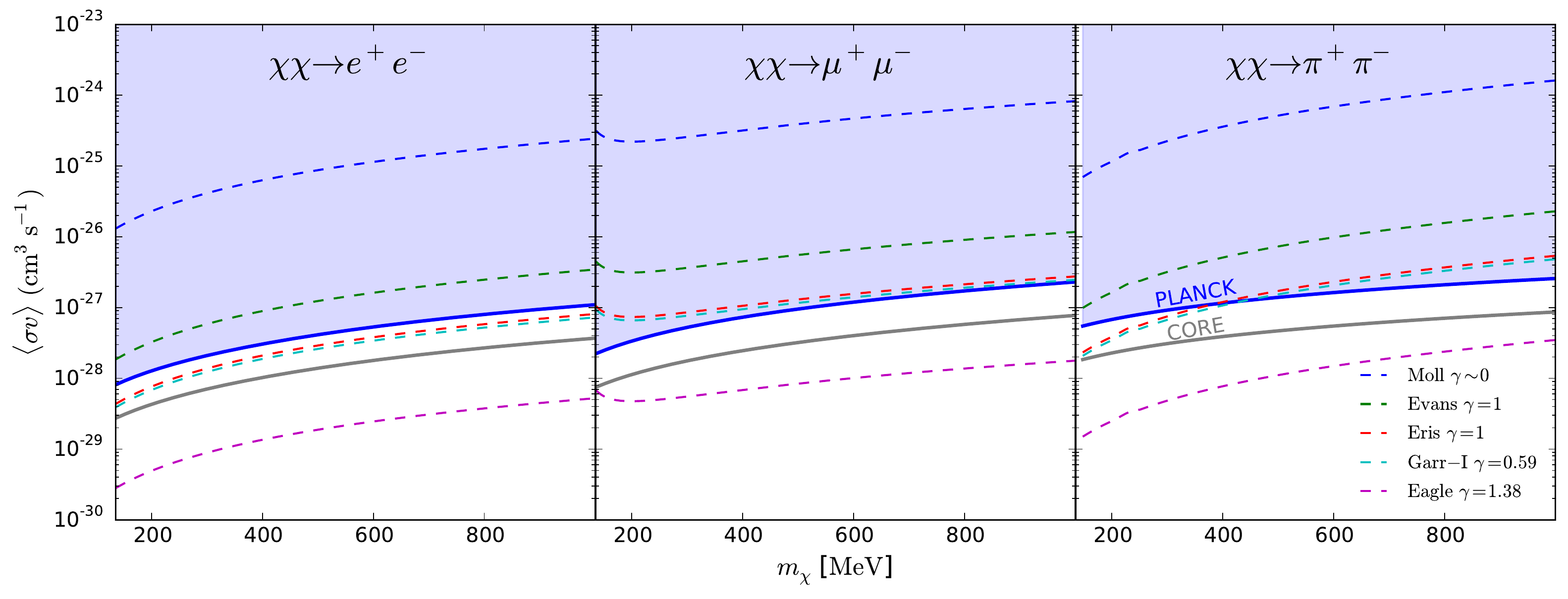}
\caption{Same as in Fig.~\ref{Fig:cases3} but for the $e^+e^-$, $\mu+\mu^-$ and $\pi^+\pi^-$ final states}
\label{Fig:cases4}
\end{figure*}

Fig.~\ref{Fig:cases3} and \ref{Fig:cases4} show  a similar analysis for the Galactic Center (GC) as target. As mentioned before in this section, we considered a broad range of possible values for the J-factor leading to different detection values for $\langle \sigma v \rangle$. The dashed lines in Fig. \ref{Fig:cases3} and \ref{Fig:cases4} represent the detection limit at $5\sigma$ taking into account different DM density profiles (different J-factors). The $\gamma$ value corresponds to the slope of the DM profile. The diffuse background for the GC was also obtained from Ref.~\cite{Strong:2004de}, and it corresponds to the Inner Galaxy (Region A) spectrum,
\begin{equation}
E^2 \frac{d\phi}{dE} \sim 1.1 \times 10^{-2} E^{0.23} \rm{cm}^{-2} \rm{s}^{-1} sr^{-1} \rm{MeV}. 
\label{eq:diffuse_GC}
\end{equation}

\section{Comparison with previous works}\label{sec:previous}

As mentioned in Sec. \ref{sec:introduction}, previous papers have addressed the topic of annihilating/decaying dark matter in the MeV energy regime, most importantly references \cite{Boddy:2015efa} and \cite{Bartels:2017dpb}. Those studies make significantly different assumptions from what presented in here in essentially all aspects of the analysis, from background, to instrumental performance, to signal intensity and spectral regions, and detectability of a signal.  In \cite{Boddy:2015efa}, for example, they considered a non-minimal dark sector, where the DM annihilates into the two final states: $\gamma\gamma$ and $\gamma\pi^0$, with the latest presenting a spectrally unambiguous signal in the hypothetical detector. Their results are largely compatible with the current CMB constraints when using Draco as the source of gamma rays. Here we considered several other annihilation final states, and we also find that  the final state $\gamma\pi^0$, would be detectable compatibly with current CMB constraints; however that will no longer be the case assuming the projected CMB constraints provided by CORE+. The GC was not considered by \cite{Boddy:2015efa} and therefore we can not compare our results with theirs on that target. 

Ref.~\cite{Bartels:2017dpb} presented an overview of the detection opportunities of MeV DM through DM annihilation, especially for  the case of leptonic final states, but they do not consider final states we take into consideration here, e.g. $\pi^+\pi^-$; in addition, their spectra for the $\mu^+\mu^-$ utilizes extrapolated results from the {\tt DarkSUSY} package well outside the range of validity of the Monte Carlo simulations used to produce the spectra (in the several GeV range). An extensive work was done in obtaining realistic projected upper  bounds by modeling the uncertainties in the diffuse background incorporated into  a Fisher forecast analysis. Additionally, Ref.~\cite{Bartels:2017dpb} includes secondary radiation from stable charged products, which they find (as we claim here) to be largely irrelevant for detection. Our results, for the GC, are quite similar to those presented in Ref.~\cite{Bartels:2017dpb}, especially for the $\gamma \gamma$, the $\gamma \pi^0$ and  $\pi^0 \pi^0$ final states, in the sense that regardless the details of the analysis, this channels seems to be quite promissory for detection, in light of current and future CMB constraints, unless the DM density profile is flat. For the case of the lepton final states, the ones more carefully studied by \cite{Bartels:2017dpb}, we find, as they do,  that detection in most of these channels would be excluded by current constraints, except in the case that the inner density profile is steeper than a standard NFW density profile. 

In summary, the analysis provided in this paper is similar to that in   \cite{Bartels:2017dpb} with a few important exceptions: i) we have considered the additional $\pi^+ \pi^-$ channel, ii) we utilize a more accurate computation of the gamma-ray production from the $\mu^+\mu^-$ final state, especially for dark matter masses close to the muon mass, iii) we have provided individual constraints for each channel arising from the assumption that the efficiency function is channel-dependent, iv) we have performed an analysis on the energy integration range $\Delta E$ to optimize the detection limit for each individual channel, resulting in more optimistic results regarding the possible detection, v) we show that the detection limit can vary broadly for different DM density profiles, and we have some cases for which the detection is possible even at the CORE precision level.

\section{Discussion}\label{sec:discussion}
In this work we have considered the indirect detection of s-wave pair annihilation of dark matter with masses in the MeV range (specifically, $m_{\pi^0} < E < 1$ GeV) with future MeV gamma-ray telescopes.  We investigated six different annihilation channels ($\gamma \gamma$, $\gamma \pi^0$, $\pi^0 \pi^0$, $e^+ e^-$ and $\mu^+ \mu^-$ and $\pi+\pi^-$), and we assumed a hypothetical detector  with specifications  similar to the proposed ASTROGAM telescope \cite{Tatischeff:2016ykb}. We then determined the optimal integration energy range for every given channel, and calculated the  values of $\langle \sigma v \rangle $ for a given mass and annihilation final state giving a $5 \sigma$ detection for the conservative case of a virtually background free target such as the Draco dSph and for the Galactic center. We then compared the required annihilation rate with the current s-wave annihilating dark matter CMB constraints, $f_{\text{eff}}\langle \sigma v \rangle m_\chi< 4.1 \times 10^{-28} \text{cm}^3 \text{GeV}^{-1}  \text{s}^{-1}$, and with future CMB constraints from COrE+, TEP, at the level of $P_{\rm{ann}}< 1.38 \times 10^{-28} \rm{cm}^{3} \rm{s}^{-1} \rm{GeV}^{-1}$.

Our main results are presented in Figs. \ref{Fig:cases1}, \ref{Fig:cases2}, \ref{Fig:cases3} and \ref{Fig:cases4}. 
For the cases of dark matter annihilating into leptons and charged pions, Fig. \ref{Fig:cases2} illustrates that  \cite{Ade:2015xua} constraints exclude the possibility of a detection from a dSph such as Draco,  Fig.~\ref{Fig:cases4} shows that for these channels, a detection is possible for the GC, but just for a selected DM density profiles. If we take the projected constraints from CORE, most of the cases for different J-factors are excluded. For the case of monochromatic photons and neutral pions, \ref{Fig:cases1} and \ref{Fig:cases3} show that a detection is generically possible and compatible with CMB constraints. 

We note that our results are generally similar to those presented in Ref.~\cite{Boddy:2015efa}, which also studied MeV dark matter candidates, with the exception that  our CMB limits were calculated with information on $f_{\rm eff}$ from {\em each individual channel}, and that we here use different assumptions for the dark matter density profile, the energy integration range, and the detector specifications. Our conclusions are, as a result of all these different choices, somewhat more optimistic than those reported in \cite{Boddy:2015efa}.

One source of uncertainty in  our analysis, as in any similar analysis, is the value of the $J$-factors, i.e. the assumed dark matter density profile. Most  of the analysis  for dwarf spheroidal galaxies report  $\log_{10}{J\hspace{1mm} \rm{GeV}^{-2} \rm{cm}^{-5}}
\approx 18.8$ \cite{Chiappo:2016xfs,Martinez:2013els} instead of the $\log_{10}(J\hspace{1mm} \rm{GeV}^{-2} \rm{cm}^{-5})\approx 19.05$ we are using.  The difference is  due to the maximum angle of integration  used to compute the factor $J$. Our analysis is sensitive to this choice since a lower value of $J$ implies stronger constraints on the  $\langle \sigma v \rangle $  vs $m_{\chi}$ plane. If the lower $J$-factor is used together with  CORE+ constraints,  this  would preclude detectability for most channels. 

On the  other side,  we  also analyzed the case for the Galactic Center, Figs. \ref{Fig:cases3} and \ref{Fig:cases4} . Given the much larger possible values for the J-factor in this case, the detection line in the $\langle \sigma v \rangle$ {\it vs} $m_\chi$ plane improves considerably, making all previously excluded channels promising for detection, even if we consider CORE+ projection constraints. The key uncertainty here is, however, the level of the background MeV emission in the Galactic center, which is largely  unknown.

Finally, the detection limits and constraints were computed assuming s-wave annihilating dark matter, and the p-wave annihilation case was not included since the CMB constraints relaxes considerably, and the prospects for gamma-ray detectability depend largely on the velocity distribution of the target dark matter distribution.\\


\section*{Acknowledgements}
We would like to thank Logan Morrison and Adam  Coogan for providing us with the code to compute the electron-positron spectrum generated by dark matter annihilations into charged pions, Francesco D'Eramo for important feedback on this manuscript, and Viviana Gammaldi for helpful discussion about the Galactic J-factor. This work was funded by a UCMEXUS-CONACYT collaborative project.  JR acknowledges financial support from CONACYT. AXGM acknowledges support from C\'atedras CONACYT DAIP research Grant No. 878/2017 and CONACYT project 182445. SP is partly supported by the US Department of Energy, grant number DE-SC0010107 .

\bibliography{references.bib}

\end{document}